\documentclass[prd,reprint, amsmath,amssymb,superscriptaddress]{revtex4-2}

\usepackage[utf8]{inputenc}
\usepackage{graphicx}
\usepackage{amsmath}
\usepackage{amssymb}
\usepackage{amsfonts}
\usepackage[inline]{enumitem}	
\usepackage{tensor}
\usepackage{braket}
\usepackage{lmodern}
\usepackage{setspace}
\usepackage[margin=2.5cm]{geometry}
\usepackage{amsthm}
\usepackage{mathtools}
\usepackage{color}
\usepackage{hyperref}
\usepackage{dsfont}
\usepackage{bm}

\DeclareMathOperator{\tr}{tr}

\renewcommand{\tr}{{\text{tr}}}

\newcommand{\ii}{\mathrm{i}}
\renewcommand{\d}{\mathrm{d}}

\newcommand{\nn}{\nonumber}

\newcommand{\bey}{\begin{eqnarray}}
\newcommand{\eey}{\end{eqnarray}}

\begin{document}

\title{Particle-field duality in QFT measurements}

\author{Maria Papageorgiou}
\email{mepapageorgiou@uwaterloo.ca}
\affiliation{Department of Physics, University of Patras, 26500 Greece}
\affiliation{Department of Applied Mathematics, University of Waterloo, Waterloo, Ontario, N2L 3G1, Canada}
\affiliation{Institute for Quantum Computing, University of Waterloo, Waterloo, Ontario, N2L 3G1, Canada}
\author{Jose de Ram\'on}
\email{jderamon@ubu.es}
\affiliation{Departamento de Fisica, Universidad de Burgos
Pza. Misael Ba\~nuelos s.n., 09001-Burgos, Spain}
\author{Charis Anastopoulos}
\email{anastop@upatras.gr}
\affiliation{Department of Physics, University of Patras, 26500 Greece}

\date{\today}

\begin{abstract}
Pointlike systems coupled to quantum fields are often employed as toy
models for measurements in quantum field theory. In this paper, we
identify the field observables recorded by such models. We show that in
models that work in the strong coupling regime, the apparatus is
correlated with smeared field amplitudes, while in models that work in
weak coupling the apparatus records particle aspects of the field, such 
as the existence of a particle-like time of arrival and resonant 
absorption. Then, we develop an improved  field-detector interaction model, adapting the formalism of Quantum Brownian motion, that is exactly solvable. This model confirms the association of field and particle properties in the strong and weak coupling regimes, respectively. Further, it can also describe the intermediate regime, in which the field-particle characteristics
`merge'. In contrast to standard perturbation techniques, this model
also recovers the relativistic Breit-Wigner resonant behavior in the
weak coupling regime.  The modulation of field-particle-duality by a
single tunable parameter is a novel feature that is, in principle,
   experimentally accessible.
\end{abstract}

\maketitle

\section{Introduction}
Quantum field theory (QFT) has proved immensely successful in describing the three fundamental interactions (weak, strong, and electromagnetic). Despite the severe challenges in its rigorous mathematical formulation, theorists have developed pragmatic tools for calculations that work really well, and provide unambiguous predictions to a large class of experiments. There is little reason to expect a failure of QFT at high energies, as long as one does not attempt to deal with gravity.

For this reason, it is really astonishing that the development of a consistent measurement theory for local measurements in QFT is an ongoing project until today.  This is not apparent in high-energy applications that focus on scattering experiments that can be treated via S-matrix theory (see discussion in \cite{Blum}). The latter conceptualizes scattering as a process with a single measurement event in the far future, so there is no reason to employ the state-update rule. This rule is employed, for example, in measurements of Bell-type correlations,  sequential measurements, and post-selected measurements. The problem is that the usual state-update rule conflicts with relativistic causality---see, \cite{AnSav22, AHS23,MD} and references therein. For this reason, even axiomatizations of QFT  \cite{Wightman, Haag} refrained from including a state-update rule into their axioms (see discussion in \cite{EV}). 

The state-update rule is essentially the rule for joint probabilities for multiple measurement events, and no probabilistic theory can work in its absence. Many experiments in quantum optics require the use of such probabilities, for example, to describe photon bunching and anti-bunching. These came to be described by photo-detection theories, the most prominent of which is that of Glauber \cite{Glauber1}.

Glauber's theory was arguably the first QFT measurement theory, and it proved remarkably successful. However, its scope is limited.  It works only for photons, and it involves approximations that may not be appropriate in set-ups of current interest, for example, quantum experiments in space that involve long-distance propagation of photons \cite{qoes, DSQL}.   These approximations can introduce friction with relativistic causality. More generally, the extension of quantum information theory to the relativistic regime, in order to take into account the effects of gravity and motion, arguably requires and consistent a practicable theory of QFT measurements \cite{AHS23, AHS23b}.


A consistent and practicable measurement theory should fulfill some requirements. A tentative and non-exhaustive list is the following: 
\begin{enumerate}
    \item {\it Causality:} Probabilities are compatible with relativistic causality, that is, there is no signaling outside the lightcone.
    \item {\it Covariance:} The probability assignment does not depend on the choice of coordinate system. However, it may depend on the state of motion of measuring apparatuses.
    \item {\it Localization:} The theory must provide a pragmatic definition of spatially localized systems that is compatible with experimental practice. This would lead to the notion of multi-partite systems and to the construction of the associated probabilities.
    \item {\it Non-relativistic limit:} The probability assignment recovers the standard non-relativistic measurement theory in an appropriate limit.
     \item {\it Practicality:} It should provide concrete predictions for experiments that cannot be modeled by S-matrix theory, for example, in quantum optics. 
\end{enumerate}

A theory fulfilling all these requirements is still missing. Progress has been made on several fronts, but a recurring feature is that when considering relativistic measurement models the apparatus performing the measurement plays a more prominent role than in the non-relativistic counterpart. 

There are two paradigms that address these challenges, with varying degrees of success. First, histories-based approaches and, second, local generalizations of scattering processes.

An example of a histories-based approach is the Quantum Temporal Probabilities (QTP) approach \cite{QTP1, QTP2, QTP3, AHS23}. This is based on the decoherent histories approach to quantum measurement that emphasizes emerging classicality in the measuring apparatus \cite{Gri, Omn1, Omn2, GeHa2, Har}. QTP emphasizes the spacetime aspects of QFT probabilities, namely, that every measurement record is localized in spacetime, and that physical probabilities must correspond to genuine densities in spacetime. 

 The second approach essentially formulates a QFT version of von Neumann's measurement theory \cite{vN, Busch} that involves switching on an interaction between the measured system and a probe in a compact region of spacetime. 
In this paradigm, sequential measurements are envisioned as sequential implementations of interactions with independent probes, all of which interact locally with the field. Therefore, these approaches deal with a generalization of particle physics-like scattering for modeling sequential measurements. When the probe is modeled as a relativistic quantum field, as in \cite{HeKr, OkOz,FeVe,FJR22}, and such that the interaction between apparatus and field satisfies some locality conditions, covariance and some versions of causality are guaranteed.


Another approach to QFT measurements that falls into the scattering-like paradigm, is the detector model approach. It is a pragmatic approach that is based on modeling the coupling of the quantum field to pointlike (or extended) quantum-mechanical systems that follow prescribed spacetime trajectories in flat or curved spacetime \cite{Perche}. Standard measurement theory is applied to the non-relativistic detector, and the induced state-update ruled for the field can be derived \cite{GGM22}. Frictions with causality and covariance arise \cite{cov,caus1,caus2,caus3} and can be handled by specifying the regime of validity of the model. 

Detector models are a simple and useful tool for exploring many issues relevant to information and measurements in QFT \cite{HLL12}. The main example of such models is the Unruh-DeWitt particle detector \cite{Unruh76, Dewitt}, originally employed in the discussion of the Unruh effect. Particle detector models, as their name suggests, were originally intended to tackle the challenges associated with the notion of \textit{particle} in relativistic quantum field theory (QFT), that is, to deal with the complications that a naive notion of particle introduces in scenarios involving non-inertial observers or curved spacetimes. Certainly, many authors argue that QFT does not admit a particle ontology \cite{Malament, Doreen, Clifton}. In addition, there is controversy around field ontology (see \cite{Kuhlmann} and references therein). In this paper, we are agnostic about this issue. We make no claim that particle detector models are fundamental, rather we view them as useful tools for extracting QFT phenomenology in certain regimes (see discussion in \cite{Clifton}). 


This paper follows the detector model approach to QFT measurements. Our original motivation was to answer the question: {\em What do detectors detect}?  Or equivalently: Which field observables are measured by pointlike detectors?   To this end, we considered two models,  one that works well in the strong coupling regime and one that is commonly used in the weak coupling regime. The first model is the von Neumann model of measurement \cite{vN} with an apparatus that consists of one localized degree of freedom that is strongly coupled to a scalar field and it behaves as a pointer variable. The second model is the Unruh-DeWitt detector coupled to the scalar field, and treated to lowest order in perturbation theory. For both models, we analyze in detail the limit in which the procedure of switching the detector on and off is irrelevant, that is, when the detector and the field interact freely without external influences. We found that in the von Neumann model, the pointer variable is correlated with spacetime averaged field amplitude, and we analyzed the exact sense in which in the Unruh-DeWitt model, the detector captures the particle aspects of the field at the weak-coupling limit.


Our results suggest that the coupling strength is a key factor in modulating field-particle duality in field-detector interactions. This property, pertaining to a fundamental aspect of QFT, could be measured in any set-up that allows for a tunable interaction strength of a localized system with a quantum field, for example, in circuit quantum electrodynamics \cite{cQED}, cavity optomechanics \cite{cop} or in analogue gravity experiments with Bose-Einstein condensates \cite{ang1, ang2}. 

For this reason, we constructed a new QFT measurement model that enables us to analyze both strong and weak coupling, but also the interpolating regimes. This model is an adaptation of the quantum Brownian motion (QBM) models, that have been extensively studied in the theory of open quantum systems \cite{CaLe, UZ, HPZ}. QBM models describe a harmonic oscillator in contact with a heat bath, itself modeled by a large number of harmonic oscillators. With some modifications that model can describe the measurement of a quantum field by an oscillator detector. 

The key point is that our model can be solved exactly, so it goes well beyond the results obtained in the lowest order in perturbation theory in the Unruh-DeWitt model, or through the drastic approximations of von Neumann's model. 
We show that, indeed,  the detector captures the field aspects at strong coupling and particle aspects, such as resonance and time-of-arrival behavior, at weak coupling.  In the regime of intermediate couplings,  field and particle characteristics "merge".  It is an advantage of solvable models that they can be used to characterize intermediate regimes, in which neither the field nor the particle concepts can be used exclusively for interpreting the detector's response. 


The structure of this paper is the following. In Sec. 2, we analyze field measurements in von Neumann's model, and in Sec. 3., we explain in what sense the Unruh-DeWitt model acts as a particle detector. In Sec. 4, we develop our QBM-based measurement model, and we analyze its properties. In Sec. 5, we summarize and discuss our results.

\section{Field measurements in von Neumann's model}
 The simplest quantum measurement model involves the recording of an observable $\hat{A}$ by a single pointer variable $\hat{X}$ with continuous spectrum. Let ${\cal H}_{S}$ be the Hilbert space of the microscopic system that is being measured and ${\cal H}_{A}$ be the Hilbert space of the apparatus. Let the Hamiltonian of the system be $\hat{H}_{S}$ and the Hamiltonian of the apparatus $\hat{H}_A$. In an ideal apparatus, the pointer variable only changes during the interaction with the microscopic system. Once this interaction is over, the pointer variable $\hat{X}$ is frozen. This means that $[\hat{H}_A, \hat{X}] = 0$.
 The essence of this analysis remains unchanged if we take $\hat{H}_A = 0$, or if we assume that $\hat{H}_A$ is negligible in comparison to the interaction Hamiltonian.

In measurement models, it is common to assume that the interaction between system and apparatus is switched on for a finite time. We consider an interaction Hamiltonian of the form
\begin{equation}
\hat{H}_{int} =  f(t)  \hat{A} \otimes \hat{P},
\end{equation}
where $\hat{P}$ is the `conjugate momentum' of the pointer variable
$\hat{X}$, i.e., $[\hat{X},\hat{P}]=i$. The switching function $f(t)$ vanishes outside the time interval $[0, T]$. 

Let $|\Omega\rangle$ be the initial state of the apparatus, and $|\psi_0\rangle$ the initial state of the measured system. It is straightforward to calculate that the probability density $p(X)$ for the pointer variable at times $t > T$
\begin{eqnarray}
p(X) = \langle \psi_0| \hat{D}^{\dagger}_X \hat{D}_X|\psi_0\rangle, \label{prob}
\end{eqnarray}
where  
\begin{align}
\hat{D}_X = \int \frac{\d k}{\sqrt{2\pi}} e^{ikX} \langle k|\Omega\rangle {\cal T} e^{-ik \int \d t f(t)\hat{A}(t)} \label{DX}
\end{align}
is the measurement operator defined in terms of the eigenvectors of $\hat{P}$ denoted as $|k\rangle$ . In Eq. (\ref{DX}), ${\cal T}$
stands for the  time-ordered expansion of the exponential and $\hat{A}(t)$ is the Heisenberg-picture operator $e^{i
\hat{H}_St} \hat{A} e^{-i \hat{H}_St}$.
It can be shown that the operator 
\begin{align}
\hat{\Pi}_X = \hat{D}^{\dagger}_X \hat{D}_X \label{PX}
\end{align}
defines a Positive-Operator-Valued measure for the time-averaged values of the observable $\hat{A}$, weighted by the function $f(t)$ \cite{AnSav08}. 

Here we apply this model to the measurement of a scalar field $\hat{\phi}$. We take $\hat{A} = \hat{\phi}(g) = \int \d^3 x \hat{\phi}({\bm x}) g({\bm x})$ for some smearing function $g({\bm x})$ that models the spatial extension of the interaction. We work in a reference frame where the detector is located at the coordinate origin, so $g({\bm x})$ is peaked around ${\bm x} = 0$.


For a free field of mass $m$, 
\begin{align}
\hat{H}_S = \frac{1}{2} \int \d^3 x \left(\hat{\pi}^2 + (\nabla \hat{\phi})^2 +m^2 \hat{\phi}^2\right). \label{HS}
\end{align}
If we denote $\hat{\phi}_t(g)= e^{i
\hat{H}_St} \hat{\phi}(g) e^{-i \hat{H}_St}$ 
The field commutation relations are 
\begin{align}
 &[\hat{\phi}_{t_1}(g), \hat{\phi}_{t_2}(g)] \nonumber \\
 &= \ii \int \d^3 x \d^3 x' \Delta(t_1-t_2, {\bm x} - {\bm x}')g({\bm x}) g({\bm x}'),
\end{align}
 where $\epsilon_{\bm p}= \sqrt{{\bm p}^2+m^2}$, and 
\begin{align}
\Delta(t, {\bf x}) =  \int \frac{\d^3p}{(2\pi)^3 \epsilon_{\bm p}} \sin({\bm p}\cdot {\bm x} - \epsilon_{\bm p} t)
\end{align} 
 is the commutator distribution for the Klein-Gordon field.



Consider a pointer variable is locally coupled to the field through the interaction Hamiltonian
\begin{eqnarray}
\hat{H}_{int} =  f(t)  \hat{\phi}(g)\otimes \hat{P}.
\end{eqnarray}
We define the spacetime smearing function $F(t, {\bm x}) = f(t) g({\bm x})$, and we write the associated smeared Heisenberg-picture field $\hat{\phi}(F) = \int \d t \d \bm{x} F(t, {\bm x}) \hat{\phi}(t,{\bm x})$. Then, we find that the 
probability density of the pointer variable after the interaction with the field is given by equation \eqref{prob} with 
\begin{align}
\hat{D}_X = w[X -  \hat{\phi}(F)],
\end{align} 
in terms of 
\begin{align}
w(x) = \int \frac{\d k}{\sqrt{2\pi}} e^{ikx - \frac{i}{2} k^2 \delta_F} \langle k | \Omega \rangle, \label{w}
\end{align}
where 
\begin{align}
\delta_F = \int \d t &\d^3x \d t' \d^3x' \theta(t - t') \nonumber\\
&\times F(t, {\bm x}) \Delta(t-t', {\bm x} - {\bm x}') F(t', {\bm x}').
\end{align}
 To derive \eqref{w} we have used the the Magnus expansion for the time-ordered exponential that appears in \eqref{DX}, namely,
\begin{align}
{\cal T} e^{-ik \int \d t f(t)\hat{\phi}_t(g)} = \exp\left[-i k \hat{\phi}(F) - \frac{i}{2}k^2 \delta_F \right]. 
\end{align}
It is straightforward to check that  $\int \d x |w(x)|^2 = 1$, hence, that Eq. (\eqref{PX})  defines a Positive-Operator-Valued-Measure (POVM). This POVM corresponds to the measurement of time-averaged smeared-field amplitudes. Indeed, we will associate the expectation value and variance of the pointer variable after the interaction to the expectation value and variance of the smeared field amplitudes.

To this end, we choose an initial state   $|\Omega\rangle$ for the pointer variables, such that $\langle \Omega|\hat{X}|\Omega\rangle = 0$ and $\langle \Omega|\hat{P}|\Omega\rangle = 0$, and no initial correlation between position and momentum. Then, $\int \d x x |w(x)|^2 = 0$, and the variance ${\cal N}^2$ of  $|w(x)|^2$ is 
\begin{align}
 {\cal N}^2  
  = \sigma_X^2  + \sigma_P^2 \delta_F^2,
\end{align}
where $\sigma^2_X$ and $\sigma^2_P$ are the position and momentum variances for  $|\Omega\rangle$, respectively. 

It follows that 
\begin{align}
\langle \hat{X}\rangle = \langle \psi_0|\hat{\phi}(F)|\psi_0\rangle, \\
(\Delta X)^2 = [\Delta \hat{\phi}(F)]^2 + {\cal N}^2,
\end{align}
  where $\psi_0$ the initial state of the field. This implies that the pointer variable $\hat{X}$ is correlated with the smeared field $\hat{\phi}(F)$. The correlation is not perfect, as it is characterized by a noise ${\cal N}$ that is intrinsic to the detector. By the uncertainty relation $
 \sigma_X \sigma_P\geq \frac{1}{2}$, we derive a lower bound for ${\cal N}$,
\begin{align}
{\cal N} \geq \sqrt{|\delta_F|}.
\end{align}
This bound does not depend on the initial state of the apparatus, but only on the localization area of the field, and the strength of the coupling as encoded in $\delta_F$. 

The signal-to-noise ratio for this system is $|\langle \hat{X}\rangle|/{\cal N}$. Hence, the necessary condition for a well-defined measurement is that
\begin{align}
|\langle \hat{\phi}(F)\rangle| >> \sqrt{|\delta_F|}\label{conddd}
\end{align}
For a state with a fixed number of particles, e.g., a wavepacket, it holds that $\langle \hat{\phi}(F)\rangle = 0$ and as a result there is no signal. This implies that this model cannot capture the particle aspects of the quantum field.

To estimate the strength of the noise in a field amplitude measurement, 
consider a Gaussian smearing function
\begin{align}
F(t, {\bf x}) = \frac{\lambda}{4\pi^2 s_X^3 s_T} e^{-\frac{{\bf x}^2}{2s_X^2} - \frac{t^2}{2s_T^2}},
\end{align}
with spatial width $s_X$ and temporal width $s_T$; $\lambda$ plays the role of the coupling constant. The function $F$ becomes proportional to a delta function for $s_X, s_T
\rightarrow 0$, while the limit of a constant interaction corresponds to $\sigma_T, \lambda \rightarrow \infty$, with $\lambda/\sigma_T$ constant.
We straightforwardly calculate
\begin{align}
\delta_F = \frac{2\lambda^2}{\pi^{7/2}} \int_0^{\infty} \frac{p^2 \d p}{\epsilon_{p}} e^{-s_X^2p^2} D(s_T\epsilon_p),
\end{align}
where $D(x) := e^{-x^2} \int_0^x dt e^{t^2}$ is the Dawson function. For $m \neq 0$, and $m s_T >> 1$, we can use the asymptotic expression for the Dawson function $D(x) \simeq \frac{1}{2x}$, to obtain $\delta_F \sim \sigma_x^{-3}\sigma_T^{-1}$,  i.e., $\delta_F$ is inversely proportional to the spacetime volume of the interaction region. For $m = 0$,
\begin{align}
\delta_F = \frac{\lambda^2}{2\pi^3} \frac{s_T}{s_X(s_X^2 + s_T^2)}.
\end{align}

 \section{The perturbative Unruh-DeWitt detector}

Next, we consider the case where the field-apparatus interaction term is much weaker compared to the Hamiltonian $\hat{H}_A$ of the apparatus. In this case, we can evaluate the detection probability using perturbation theory.   The Unruh-DeWitt detectors have been used extensively as measurement models in this regime.  The weak coupling approximation is physically relevant, as it applies, for example to photodetection and to neutrino detection.

Let us denote by $|\epsilon\rangle$, the eigenstates of the Hamiltonian $\hat{H}_A$, $\hat{H}_A |\epsilon\rangle = \epsilon |\epsilon\rangle$. We take the initial state of the apparatus $|\Omega\rangle$ to coincide with the ground state of $\hat{H}_A$, and we choose the energy scale so that $\hat{H}_A |\Omega\rangle = 0$. We assume an interaction Hamiltonian of the form $f(t) \hat{\mu} \hat{\phi}(g)$, as in Sec. 2, where $\hat{\mu}$ is an operator on the Hilbert space ${\cal H}_A$ of the apparatus. The switching function $f(t)$ vanishes for $t$ outside $[0, T]$.

To leading order in the interaction, the excitation probability for the apparatus is
\bey
P = \sum_{\epsilon > 0} |\mu(\epsilon)|^2 \int \d t\d^3x \int \d t' \d^3x' F(t, {\bm x}) F(t, {\bm x}') \nonumber \\ \times e^{-i \epsilon (t - t')} \langle \psi_0|\hat{\phi}(t',{\bm x}')\hat{\phi}(t, {\bm x})|\psi_0\rangle, \; \; \label{detprob}
\eey
where $\mu(\epsilon) = \langle \epsilon|\hat{\mu}|\Omega\rangle$, and again $F(t, {\bm x}) = f(t) g({\bm x})$.

Consider a field state  $|\psi_0\rangle$ with a definite number of particles. Then,
$P$ in Eq. (\ref{detprob}) is a sum of two terms, $P = P_0 + P_1$. The first term
\bey
P_0 = \sum_{\epsilon > 0} |\mu(\epsilon)|^2 \int \frac{\d \bm{k}}{ 2\epsilon_{\bm k}} |\tilde{F}(\epsilon + \epsilon_{\bm k}, {\bm k})|^2 \label{P0}
\eey
is independent of the initial state of the quantum field, and it is due to the excitation of the vacuum from the switching of the interaction.  In Eq. (\ref{P0}), $\tilde{F}(\omega, {\bm k})$ is the Fourier transform of $F(t, {\bm x})$, and $\d{\bm k} = \d^3k/(2\pi)^{3/2}$.  $P_0$ functions as a noise term in the measurement scheme; it becomes stronger for smearing functions that are strongly localized in spacetime. If $\tau$ is the effective duration of the interaction, the noise is suppressed if $\tau \epsilon_1 >> 1$, where $\epsilon_1$ is the energy of the first excited state in the detector. 

The second term $P_1$ depends on the one-particle reduced density matrix of the field $\rho_0({\bm k},{\bm k}') = \langle \psi_0|\hat{a}^{\dagger}_{\bm k} \hat{a}_{\bm k}|\psi_0\rangle$. For simplicity, we consider a pure $\rho_0({\bm k}, {\bm k}') = \psi_0({\bm k}) \psi_0^*({\bm k}')$, to obtain
\bey
P_1 &= \sum_{\epsilon >0} |\mu(\epsilon)|^2  \left| \int \frac{\d \bm{k}}{\sqrt{2\epsilon_{\bm k}}}  \tilde F^*(\epsilon_{\bm k}+\epsilon, \bm{k})\psi_0({\bm k})\right|^2\nonumber \\
   &+ \sum_{\epsilon > 0}|\mu(\epsilon)|^2 \left|\int \frac{\d \bm{k}}{\sqrt{2\epsilon_{\bm k}}}  \tilde F^*(\epsilon_{\bm k}-\epsilon, \bm{k}) \psi_0(\bm{k})\right|^2. \label{P1}
\eey
is the "signal" as it depends on the initial field state,  e.g, a wavepacket of the form
\begin{equation}
    \ket{\psi_0}= \int \d \bm{k} \psi_0(\bm{k}) \hat{a}^{\dagger} (\bm{k}) \ket{0}.
\end{equation} 
A proper measurement requires a large signal-to-noise ratio, hence, $\epsilon_1 \tau >> 1$. In this regime, the first term in the right hand side of Eq. (\ref{P1}) is negligible. 

This model functions as a particle detector for sufficiently large duration of the interaction. To show this, we write the excitation probability (\ref{P1}) as a spacetime integral:
\begin{align}
    &P_1= \sum_{\epsilon > 0} |\mu(\epsilon)|^2 \left|\int \d t \d^3 x e^{i\epsilon t}F(t, \bm x)\psi_{\textsc{nw}}(t,\bm x)\right|^2 \nonumber \\
    &+  \sum_{\epsilon > 0} |\mu(\epsilon)|^2 \left|\int \d t \d^3 x e^{-i\epsilon t} F(t,\bm{x})\psi_{\textsc{nw}}(t,\bm x)\right|^2 \nonumber
\end{align}
where we   defined the Newton-Wigner wavefunction \cite{NW}
 \begin{equation}
\psi_{\textsc{NW}}(t,\bm {x}) = \int \frac{\d\bm{k}}{\sqrt{2 \epsilon_{\bm{k}}}}\psi_0(\bm{k}) e^{i \bm{k}\bm{x} - i \epsilon_{\bm{k}} t}.
\end{equation}
 
Assume an almost monochromatic initial state with momentum ${\bm p}_0$, such that $\psi_{\textsc{NW}}(0, {\bm x})$ is concentrated around a point ${\bm x}_0$. Then, $\psi_{NW}(t,{\bm x})$ is concentrated in a world tube $W$ that surrounds surrounding the classical path ${\bm x}(t) = {\bm x}_0 + {\bm p}_0 t$. Let $C$ stand for the region of support of the function $F$. The probability density $P_1$ is appreciably different from zero only if $W \cap C \neq \emptyset$. In a semi-classical language, this means that the bundle of classical particle paths associated with the initial conditions cross into the interaction region. This conforms to our classical intuitions about the behavior of a particle detector.  

Realistic field-detector couplings are typically {\em isotropic}, that is, the smearing function $g({\bm x})$ is a function only of $|{\bm x}|$. In this case, $\tilde{F}$ depends only on $k = |{\bm k}|$, and the excitation probability $P_1$ involves an integral $\int d {\bf n} \psi(k, {\bf n})$, where ${\bf n} = {\bf k}/|{\bf k}|$. Writing $\psi({\bm k}) = \sum_{\ell,m_{\ell}} \psi_{\ell, m_{\ell}}(k) Y_{\ell, m_{\ell}}({\bf n})$, in terms of spherical harmonics $Y_{\ell, m_{\ell}}({\bf n})$, we find that only the zero angular momentum $\psi_0,0(k)$ component  survives. This means that propagation is essentially one-dimensional, along the axis that connects the source and the detector, again in accordance with our classical intuition.  The fact that the detector `clicks' in response to zero angular momentum spherical waves (and not plane waves) can be used to explain the counter-intuitive behavior that was reported in \cite{ET}, that in certain spacetime dimensions the detector's excitation probability decreases in the monochromatic limit. 

In the limit of an interaction that is adiabatically switched on, that is, for $f(t) = 1$, we have $\tilde{F}(\epsilon, {\bf k}) = 2\pi \delta(\epsilon) g(k)$, and 
\bey
P_1 = \sum_{\epsilon > 0} \frac{\epsilon |\mu(\epsilon)g(\sqrt{\epsilon^2-m^2})|^2 }{8 \pi^2 (\epsilon^2 - m^2)} |\psi_{0,0}(\sqrt{\epsilon^2 - m^2})|^2. \nonumber
\eey
If the energy levels of the detector are well separated, and the initial state is approximately monochromatic,  then the detection probability is characterized by resonances with respect to the incoming energy. However, these resonances are not of the standard form (Breit-Wigner for scattering, or Wigner-Weisskopf for photodetection), but they mirror the profile of the wavefunction $\psi_{0,0}$. 
This behavior is arguably unphysical, or at least, it does not account for the observed resonances. However, this is to be expected, because resonances cannot be consistently identified from the leading-order terms in perturbation theory \cite{RoTh, CoTa}; the derivation of the standard form requires at least a partial resummation of a perturbative series---see, for example, chapter 7 in \cite{PeSc}, or chapter 18 in \cite{Ana23}.

\section{Quantum Brownian motion as a QFT measurement model}
In Secs. 2 and 3, we saw that a field-particle interaction with an UDW-type coupling corresponds (i) to a measurement of field properties if the apparatus Hamiltonian is negligible in compared to the interaction term, and (ii) to a measurement of particle properties if the interaction is weak compared to the apparatus Hamiltonian. In this section, we present a model that incorporates both cases for different parameter regimes. Furthermore, this model does not require the  trick of switching on the field-apparatus interaction.

To this end, we consider a Hamiltonian $\hat{H} = \hat{H}_S + \hat{H}_A + \hat{H}_I$, where $\hat{H}_S$ is the field Hamiltonian (\ref{HS}), the apparatus Hamiltonian 
\begin{align}
\hat{H}_A = \frac{1}{2M}\hat{P}^2 + \frac{1}{2}M \omega_0^2 \hat{X}^2
\end{align}
describes an harmonic oscillator of mass $M$ and frequency $\omega_0$, and the interaction Hamiltonian is $\hat{H}_I = \lambda \hat{\phi}(g)\hat{X}$.

The total Hamiltonian is of the Quantum Brownian motion (QBM) form
\begin{align}
\hat{H} = \frac{\hat{P}^2}{2M}  + \frac{1}{2}M\omega_0^2 \hat{X}^2 &+ \sum_i\left( \frac{\hat{p}_i^2}{2m_i}+ \frac{1}{2}m_i\omega_i^2 \hat{q}_i^2 \right) \nonumber\\
&+ \hat{X} \sum_i c_i \hat{q}_i,
\end{align}
i.e., it describes the interaction of a distinguished oscillator with a bath of oscillators labeled by $i$. 

QBM models admit  exact solutions for their time evolution \cite{HPZ, HaYu, AKM, FRH, UZ}. Here, we follow the presentation of \cite{Ana23}. We first define the dissipation kernel
\bey
\gamma(t) := \sum_i \frac{c_i^2}{2m_i \omega_i^2} \cos(\omega_i t),
\eey
and then we identify the function $u(t)$ that solves the integrodifferential equation
\bey
\ddot{u}(t) + \bar{\omega}^2 u(t) + \frac{2}{M} \int_0^t dt' \gamma(t-t') \dot{u}(t') = 0,\label{udefeq2}
\eey
subject to initial conditions $u(0) = 0 $ and $\dot{u}(0) = 1$. We have defined the regularized frequency $\bar{\omega}$ by $\bar{\omega}^2 = \omega_0^2 - 2\gamma(0)/M$.

Then, we can solve for $\hat{X}(t)$ and $\hat{P}(t)$ in the Heisenberg picture,
\bey
\hat{X}(t) &=& \dot{u}(t) \hat{X}(0) + \frac{1}{M}u(t) \hat{P}(0) \nonumber \\ 
&-&\frac{1}{M} \int_0^t dt' u(t - t') \sum_i c_{i} \hat{q}_i^0(t'), \label{Xsol}
\\
\hat{P}(t) &=&  M \frac{d}{dt} \hat{X}(t).
\eey
where $\hat{q}_i^0(t)$ stands for the evolution of the bath harmonic oscillators in the absence of interactions. 

A QBM model is specified by the distribution of frequencies $\omega_i$ and couplings $c_i$ among the environment oscillators. These determine the dissipation kernel uniquely, and then, the function $u(t)$, by the solution of Eq. (\ref{udefeq2}). In the context of the field-detector interaction, 
the bath oscillators are labeled by the momenta ${\bf k}$, $\omega_i$ corresponds to $\epsilon_{\bf k}$, the oscillator masses $m_i = 1$, and the coupling constants $c_i$ correspond to 
$\lambda \tilde{g}({\bf k})$, where $\tilde{g}$ is the Fourier transform of the smearing function $g({\bf x})$. The dissipation kernel is then,
\bey
\gamma(t) = \frac{\lambda^2}{4\pi^2} \int_0^{\infty}  \frac{dk k^2|\tilde{g}(k)^2|}{\epsilon_k^2}\cos(\epsilon_kt),
\eey
while the Heisenberg-picture operators
$\hat{X}(t)$ and $\hat{P}(t)$ are given by
\begin{align}
    \nn\hat{X}(t) =& \dot{u}(t) \hat{X}(0) + \frac{1}{M}u(t) \hat{P}(0) \\
    &-\frac{\lambda}{M} \int_0^t dt' u(t - t') \hat{\phi}_{g}(t'), \label{Xsol}
\\
\nn\hat{P}(t) = & M \ddot{u}(t) \hat{X}(0) + \dot{u}(t) \hat{P}(0) \\&- \lambda\int_0^t dt' \dot{u}(t - t') \hat{\phi}_{g}(t'). \label{Psol}
\end{align}
Here, $\hat{\phi}_{g}(t)$ is the Heisenberg-evolution of the smeared field $\hat{\phi}(g)$.

We assume that the apparatus oscillator is prepared in an initial state with $\langle \hat{X} \rangle = 0$, $\langle \hat{P}\rangle = 0$ and $c_{XP} = 0$. Then, we obtain the following equations for the mean and the variances of the detector's position and momentum.
\begin{align}
  & \langle \hat{X}(t)\rangle = -\frac{\lambda}{M} \int_0^t dt' u(t-t') \langle \hat{\phi}_g(t')\rangle,\\ 
  &\langle \hat{P} (t)\rangle = - \lambda \int_0^t dt' \dot{u}(t-t') \langle\hat{\phi}_g(t')\rangle,  \label{expPt}\\
  &\langle \hat{X}^2(t)\rangle =  \sigma_X^2\dot{u}^2(t)+ \frac{\sigma_P^2}{M^2}{u}^2(t) \\
&+ \frac{\lambda^2}{M^2} \int_0^t dt' \int_0^t dt'' u(t-t') u(t-t'')\langle \hat{\phi}_g(t')\hat{\phi}_g(t'')\rangle, \nn\label{deltaX}\\
&\langle \hat{P}^2(t)\rangle = M^2\sigma_X^2\ddot{u}^2(t)+ \sigma_P^2\dot{u}^2(t)\\
&+ \lambda^2 \int_0^t dt' \int_0^t dt'' \dot u(t-t') \dot u(t-t'')\langle \hat{\phi}_g(t')\hat{\phi}_g(t'')\rangle\nn.
\end{align}

To proceed further, we need to evaluate the function $u(t)$. We will analyze the case $m = 0$. For concreteness, we take $|g(k)|^2 = e^{-k/\Lambda}$, where $\Lambda$ is an ultra-violet cut-off; $\Lambda^{-1}$ is the size of the interaction region. Then, 
\bey
\gamma(t) = \frac{\lambda^2}{4\pi^2} \frac{\Lambda}{1+\Lambda^2 t^2}.
\eey
For $\Lambda >> \bar{\omega}$, we can approximate $\gamma(t)$ in Eq. with $\frac{\lambda^2}{4\pi} \delta (t)$. Then,  Eq. (\ref{udefeq2}) becomes
\begin{align}
\ddot{u} + 2\Gamma \dot{u} + \bar{\omega}^2 u  = 0,
\end{align}
where $\Gamma = \frac{\lambda^2}{16 \pi M}$. This is the equation of a  damped harmonic oscillator, with solution
\bey
u(t) = \left\{ \begin{array}{ccc} \omega^{-1} \sin(\omega t)e^{-\Gamma t},& \omega = \sqrt{\bar{\omega}^2- \Gamma^2}, &  \Gamma <  \bar{\omega} \\
w^{-1} \sinh(w t) e^{-\Gamma t}, &  w = \sqrt{\Gamma^2 - \bar{\omega}^2},& \Gamma >  \bar{\omega}
\end{array} \right.  \nonumber
\eey

 At the strong-coupling limit, $\bar{\omega}/\Gamma << 1$, and $u(t) \simeq \frac{1}{2\Gamma}(e^{-\frac{\bar{\omega}^2}{2\Gamma}t} - e^{-2\Gamma t}) $, i.e., $u(t)$ interpolates between 0 at $t = 0$ and $\frac{1}{2\Gamma}$ at times $t$, such that $\Gamma^{-1} << t << \Gamma \bar{\omega}^{-2}$. Therefore, we can approximate $u(t) \simeq \frac{1}{2\Gamma}\theta(t)$. Then, Eq. (\ref{expPt}) becomes,
 \bey
\langle \hat{P} (t)\rangle = - \frac{\lambda}{4\Gamma} \langle \hat{\phi}_g(t)\rangle, \label{scP}
 \eey
that is, the detector's momentum observable records the field's expectation value. Eq. (\ref{scP} holds for any field pulse that (i) arrives at the detector at times $t >> \Gamma^{-1}$, and (ii) is of duration $\tau$, such that $\bar{\omega}^2 \tau/\Gamma<< 1$. In this regime, $(\Delta P)^2(t) \simeq \frac{\lambda^2}{16 \Gamma^2} (\Delta \phi_g)^2(t)$. Hence, the oscillator acts as an antenna: it records faithfully "classical" field states, that is, field states with small quantum fluctuations. 

For an initial state with a fixed number of particles, $\langle \hat{X}(t) \rangle = 0$ and $\langle \hat{P}(t) \rangle= 0$. Then, the natural quantity for a detection signal is the energy of the harmonic oscillator,
\bey
 \langle \hat{H}(t) \rangle = \frac{(\Delta \hat{P})^2(t)}{2M} +\frac{M\bar{\omega}^2(\Delta \hat{X})^2(t)}{2}. \label{hamho}
\eey
The expressions simplify if we assume an initial state that is localized at ${\bf x} = {\bf L}$, such that $|{\bf L}| \Gamma >> 1$. Then, the signal from incoming particles appears at times $t> |{\bf L}|$, when many transient noise terms have vanished. We obtain (for $m = 0$),
\bey
 \langle \hat{H}(t) \rangle = h(t) +{\cal N},
\eey
where 
\bey
{\cal N} = \frac{2 \Gamma}{\pi} \int_0^{\infty} \frac{dk k |g(k)|^2 (\bar{\omega}^2 + k^2)}{(\bar{\omega}^2 - k^2)^2 + 4 \Gamma^2 k^2}
\eey
 is the vacuum noise, and 
 \begin{align}
h&(t)\\
&= 8 \pi \Gamma  \left| \int \frac{d{\bf k}}{\sqrt{2\epsilon_{\bf k}}} \psi_0({\bf k}) g({\bf k}) \frac{\bar{\omega}^2 + \epsilon_{\bf k}^2}{\bar{\omega}^2 + (\Gamma - i \epsilon_{\bf k})^2} e^{-i \epsilon_{\bf k}t}\right|^2 \nonumber
 \end{align}
is the detection signal\footnote{ The noise term diverges logarithmically as the size of the detector vanishes, that is, $\tilde g\to1$. Note that we have used the pointlike renormalized dynamics for solving the integrodifferential equation, but we kept the smearing in the source term in order to calculate the noise. Although this might seem artificial, it can be shown that this noise term represents the leading order behavior of the noise in a power expansion involving a UV cutoff.}. We assume that the initial single-particle state is almost monochromatic at ${\bf k} = {\bf k}_0$. Then, 
 \bey
h(t) = \frac{8 \pi \Gamma |g(\epsilon_0)|^2 (\epsilon_0^2 +\bar{\omega}^2)}{(\bar{\omega} - \epsilon_0^2)^2 + 4 \Gamma^2 \epsilon_0^2} |\psi_{NW}(t, 0)|^2, 
\eey
where $\epsilon_0 = |{\bf k}_0|$.

We see that the detection signal appears after the time $t = |{\bf L}|$ that the Newton-Wigner wave packet arrives at ${\bf x} = 0$, as in the perturbative Unruh-DeWitt model of Sec. 3. Note that no switching function has been employed. Except for noise terms, the field interacts with the detector only when the wavepacket has reached the locus of the detector.

We also note that for $\Gamma << \bar{\omega}$, the signal is characterized by a resonance that is  independent of the shape of the initial state. The resonance is described by the relativistic Breit-Wigner distribution \cite{BreitWigner}. 

We conclude that the harmonic oscillator in this model functions as a particle detector that records energy at small $\Gamma$, and as an antenna that records the field's amplitude at large $\Gamma$. The model is well defined also in the regime of intermediate $\Gamma$, thus interpolating between the field regime (also described by von Neumann's model) and the particle regime (also described by the perturbative Unruh-DeWitt detector).

\section{Conclusions}
In this article, we explored the principle of field-particle duality using simple models for QFT measurements. We saw that von Neumann's theory describes the measurement of classical field amplitude in the regime of strong field-detector coupling. We also saw that the perturbative Unruh-DeWitt model indeed describes particle detection in the regime of weak field-detector coupling, with the caveat that it does not provide a description of resonances reproducing observed experimental results. Indeed, resonances of the Breit-Wigner type require absorption effects that can only be captured nonperturbatively, or after a partial resummation of the perturbative series. 

Then, we presented a new measurement model that is based on QBM. This model is exactly solvable. It enables us to show that in the long-time limit, field amplitude measurements occur at strong coupling and particle measurements at weak coupling. It also accounts for the regime of intermediate couplings that corresponds to the transition from field to particle behavior. The fact that this transition is modulated by a single parameter is a novel aspect of field-particle duality, which we expect to be experimentally accessible in set-ups that allow for tunable coupling of a quantum oscillator with a quantum field. 


Besides experimental relevance, the model presented here is a substantial improvement over existing models for QFT measurements that are based on the coupling of fields to pointlike systems. Being exactly solvable, it provides a consistent analysis of the strong coupling regime, and it also provides a consistent description of resonances. Furthermore, the model remains exactly solvable if we introduce multiple detectors \cite{AKM, FRH}, for example, in order to analyze causality in relation to field measurements, relativistic quantum communication \cite{RC}, or even measurements in an extended spatial region.

\textbf{Acknowledgements:} MP acknowledges support of the ID$\#$ 62312 grant from the John Templeton Foundation, as part of the ‘The Quantum Information Structure of Spacetime’ Project (QISS). JRR's work was supported by the Regional Government of Castilla y León (Junta de Castilla y León) and by the Ministry of Science and Innovation MICIN and the European Union NextGenerationEU (PRTR C17.I1). JRR would like to thank the Univerisity of Patras and Charis Anastopoulos' group for the hospitality during the course of this research.  

\end{document}